\documentclass[12pt,fleqn]{article}

\usepackage{latexsym}
\usepackage{epsfig}
\usepackage{amsmath}
\usepackage[boxed]{algorithm2e}

\newcommand{\enproof} {\hfill $\Box$ \bigskip}

\newtheorem{theorem}{{\bf Theorem}}[section]
\newtheorem{lemma}{{\bf Lemma}}[section]

\begin{document}

\title{A Provably Linear Time, In-place and Stable Merge Algorithm via the Perfect Shuffle
\thanks{This work was supported in part by the Natural Sciences and Engineering
        Research Council of Canada}}
\author{{\sc John Ellis$^\dagger$} and { \sc Ulrike Stege$^\dagger$}  \\
\ \ \\
     {\small \em $^\dagger$Department of Computer Science} \\
     {\small \em University of Victoria, P.O. Box 3055} \\
     {\small \em Victoria, British Columbia, V8W 3P6, Canada}}  
\date{\today}

\maketitle
\begin{abstract}
We reconsider a previously published (Dalkilic et al.) algorithm for merging lists
by way of the perfect shuffle permutation.
The original publication gave only experimental results which, although consistent
with linear execution time on the samples tested, provided no analysis.
Here we prove that the time complexity, in the average case, is indeed linear, although there is 
a $\Theta(n^2)$ worst case.
This is then the first provably linear time merge algorithm based on the use of the perfect shuffle.
We provide a proof of correctness, extend the algorithm to the general case where the lists are of 
unequal length and show how it can be made stable, all aspects not included in the original presentation
and we give a much more concise definition of the algorithm.
\end{abstract}
\noindent
\textbf{keywords: }{algorithms, perfect shuffle, merging, sorting, stability.}
\newpage
\section{Introduction}
In the context of data processing, to merge lists is to create one list, sorted 
on some key, from the elements of two sorted lists.
The standard merge algorithm is simple and uses only time linear in the input size
but it duplicates the input space.
Merge algorithms that use no extra space other than the program variables and 
possibly that space used for recursion are sometimes called \emph{in-place}.
Another significant property, besides time and space complexity, is ``stability'',
which means that the order of equal elements is guaranteed to be preserved.
A linear time, in-place, stable merge algorithm can be used to realise an in-place,
stable sorting algorithm with optimal time complexity $O(n \log n)$ by repeated
merging.
If one is using the merge to sort on more than one key, stability is essential.

The problem of constructing an algorithm with the three desirable properties:
linear time, using $O(\log^2n)$ space and stability, has a long history, having 
been first posed in \cite[Chapter 5, Section 5, Exercise 3]{Knu:1973:ACP}.
Since that time several solutions have been proposed.
Partial surveys of this work are given in \cite{EllMar00,Dalkilic:2013}.

Most solutions use block rotation techniques where segments of the arrays used
to represent the lists are cyclically shifted.
A new approach was introduced in \cite{EllMar00} which begins the merge by 
performing a ``perfect shuffle'' on equal length sub-lists of the input, before
any comparisons are made.
A perfect shuffle intersperses the elements of equal length sub-lists so that the order
of elements in each list is preserved and every other element in the result comes
from the same input list.
A perfectly shuffled list is of even length.
A \emph{2-ordered} list has the same properties except that it is not necessarily of
even length.

The list resulting from the shuffle is not necessarily sorted but one might expect that,
on the average, elements will already be close to their correct position, after
the single shuffle operation.
It remains only to tidy up the shuffled list.

In \cite{EllMar00}, the tidying is accomplished by breaking the shuffled list
into a sequence of \emph{d-strings}, where a \emph{d-string} is a maximal length
sub-list whose first element is greater than the last.
The d-strings are then unshuffled and adjacent unshuffled segments are recursively
merged.
The time complexity analysis given showed only time $O(n \log \log n)$ and that 
not on the given algorithm but only on an elaboration thereof.
The reported results of experiments were not consistent with linear time.

In \cite{Dalkilic:2013} another way to do the tidying was described.
This process, described again below, moves from left to right (or \emph{vice versa})
taking items from what remains of the shuffled list and adding them, via a rotation,
to an already sorted list.
This method is an improvement because it is relatively simple to implement and
does not need recursion, which can bring the space complexity down to the absolute minimum
of $O(\log n)$, i.e., only a constant number of variables is used.
Most importantly, experiments gave time complexity results consistent with linear time.
However the authors of \cite{Dalkilic:2013} did not give an analysis of the time complexity,
only the experimental results.

It is the purpose this paper to prove that the time complexity of the new algorithm \cite{Dalkilic:2013} is,
in the average case, indeed linear, although there is a worst case requiring $\Theta(n^2)$ time.
We also present the algorithm in a more concise form which we think helps to
clarify its structure and the analysis.
We also provide the general form of the algorithm that covers lists of unequal length,
which was missing from the original presentation, and we update a review of shuffling
methods, which are essential to the simplicity and practicality of the complete process.
Finally, we remind readers that stability, if required, falls out very naturally from 
the shuffling process and show how it can be realised in this new algorithm.

Section 2 defines the algorithm  and gives a proof of correctness.  The proof of average case,
linear time complexity is given in Section 3.  
In Section 4 we list some of the more promising Shuffling methods so far proposed, one of which
was not mentioned in \cite{Dalkilic:2013}.
Section 5 shows how to achieve stability, should it be required.

\section {The Algorithm}
Suppose the two sorted lists to be merged are represented by contiguous segments of an array $A$.
We also assume that there are no equal elements.
The case where there may exist equal elements and stability is required is covered in Section 5.
We first define the core of the process, in Section 2.1,
by assuming that the lists to be merged are of equal length, say $N/2$.
We generalise to lists of unequal length in Section 2.3.

\subsection{The case of equal length lists}
\label{algo}
\begin{figure}
\begin{center}
\epsfig{file=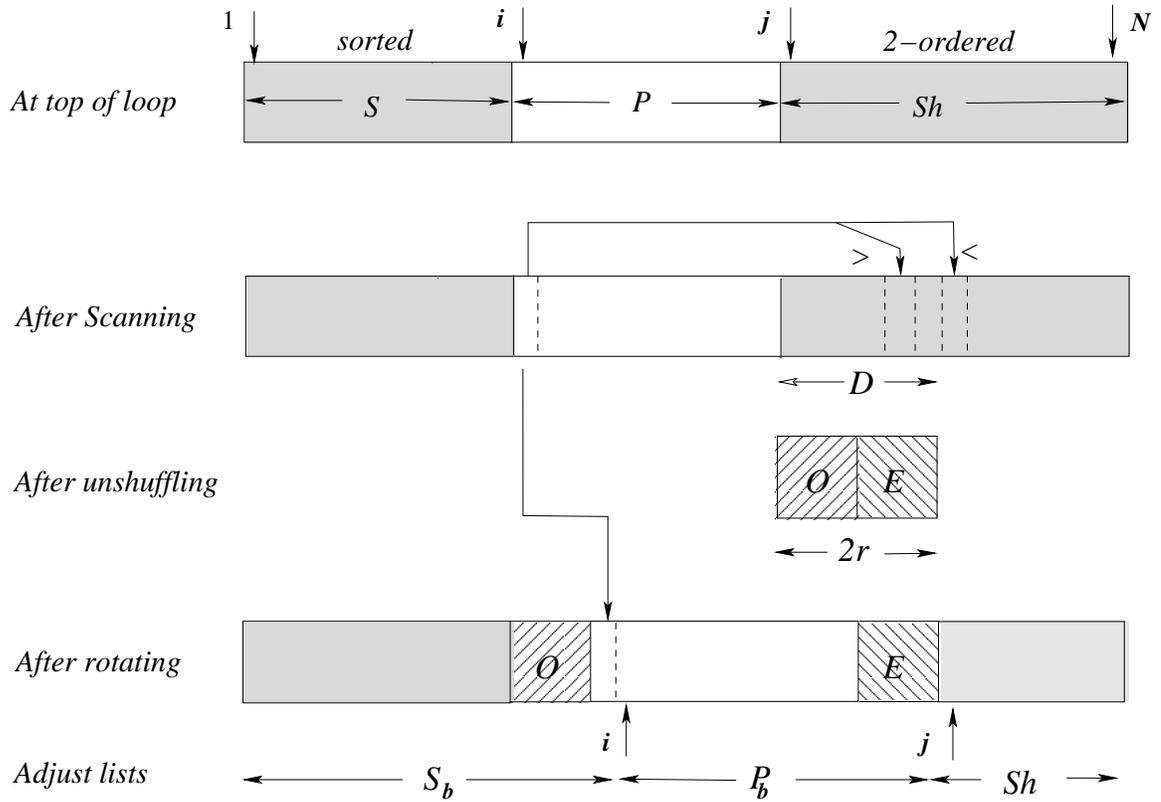, width=6in}
\end{center}
\caption{The three sub-arrays}
\label{sublist1}
\end{figure}

Suppose the lists are of equal length.
The process maintains three lists:
a sorted list, $S$, an intermediate  list, $P$ and a 2-ordered list, $Sh$.
Figure \ref{sublist1} illustrates these structures.
The array indices $i$ and $j$ are used to delineate the extents of the three lists.
Index $i$ defines the beginning of $P$ and $j$ the beginning of $Sh$. 
The list $S$ comprises $A[1]$ - $A[i-1]$, $P$ is $A[i]$ - $A[j-1]$ and $Sh$ is $A[j]$ - $A[N]$.

The algorithm Right-going-merge, see Algorithm 1, uses four procedures.
The procedure Scan returns an integer $r$ such that $A[j]$ - $A[j+2r-1]$ is a maximal, even length prefix of 
$Sh$, denoted $D$, such that all odd indexed elements are less than $A[i]$, the first element of $P$.
Scan is only invoked if $|Sh| \geq 2$.
The procedure Shuffle performs an in-shuffle on the input lists, assumed to be of equal length,
i.e., only the interior elements are moved, the first and last elements are left unmoved.

The procedure Unshuffle performs the inverse of Shuffle, i.e., an un-in-shuffle, on $D$ to produce the two lists 
$O$ and $E$. 
Shuffling methods are discussed in Section 4.
The procedure Rotate circularly shifts the two adjacent segments of $A$ that represent $P$ and $O$
to the right by $r$.
See Figure \ref{sublist1}.
We call this procedure \emph{right-going-merge} because the scan proceeds from left to right.
As described in Section 2.3, to handle the case where the lists are not of equal length, we
also use the mirror image of this procedure, called the \emph{left-going-merge}, which scans
from right to left.

\begin{algorithm}
\begin{tabbing}
1234\=1234\=1234\=1234\=12345\=12345\=12345\=12345         \kill

\>Create $Sh$ by applying Shuffle to the two, equal length lists.\\
\>\{Recall that $A[i]$ is $P[1]$ and $A[j]$ is $Sh[1]$\}\\
\>$i :=$ index of first element in $Sh$;  $j := i+1$;\\
  \\
\>\textbf{while} \textbf{not} $Sh$ is empty \textbf{do}\\ 
\>\>\textbf{if} $P[1] < Sh[1]$ \textbf{then} \{adjust lists\} i := i+1\\
\>\>\>\textbf{if} $|P| = 0$ \textbf{then} $j := j + 1$; complement(\emph{type}) \textbf{fi}\\
\>\>\textbf{else if} $|Sh| = 1$ \textbf{then} $r$ := 1; Rotate; \{adjust lists\} i := i+1; j:=j+1 \\
\>\>\textbf{else} \{Figure 1\} Scan; Unshuffle; Rotate; \\
\>\>\>\{Adjust lists\} $i := i+r+1;$  $j := j+2r$ \textbf{fi} \textbf{fi}\\
\>\textbf{endwhile};\\
\end{tabbing}
\caption{Right-going-merge}
\end{algorithm}

\subsection{Correctness on equal length lists}

We call the input list occupying the lower indexed part of the array the left list and the 
other the right list.
We define the \emph{type} of an element to be either \emph{left} or \emph{right}
depending on which of the input lists it was a member.

The process maintains the following properties of the three lists which constitute a loop
invariant assertion:
\begin{description}
\item{A0:} $P$ is not empty.
\item{A1:} the elements in $S$ are in order and there is no element in $S$ greater than any element in 
$P$ or in $Sh$.
\item{A2:} the elements in $P$ are all of the same type and are ordered.
\item{A3:} $Sh$ is \emph{2-ordered} and such that all odd elements are of one type and
all the even elements are of the other type.
\item{A4:} all elements in $P$ are less than any element of the same type in $Sh$.
\item{A5:} the first elements in $P$ and in $Sh$ are of different types.
\end{description}

We use the convention that $S[k], P[k]$ and  $Sh[k]$ refer to the $k^{th}$ element in the
respective list, not the $k^{th}$ element in $A$.
The initial Shuffle and definitions of $i$ and $j$, before the loop, leave $S$ empty, 
$P$ with what was the first element of $Sh$ and $Sh$ 2-ordered.
So $A0$ - $A5$ hold at first entry to the loop.
When the loop is exited $Sh$ is empty, since the loop condition is then false.
Then, by $A1$ and $A2$, $A[1]$ through $A[N]$ is ordered.
It remains to show that if $A0$ through $A5$ hold at the top of the loop they still
hold at the bottom.

The body of the loop includes three sets of actions, chosen by the \textbf{if} statements.
At entry to the loop we know that $Sh$ is not empty, by the loop condition, and $P$ is not
empty, by $A0$.
At the first \textbf{if}, if
$P[1] < Sh[1]$ then, by $A1$ - $A5$, $P[1]$ is the next smallest element and should 
be added to $S$, which is accomplished by incrementing $i$. 
That shortens $P$. If $P$ is now empty then $j:= j+1$ restores $A0$, $A2$, $A4$ and $A5$ because
of $A3$, otherwise $A0$ - $A5$ are unaffected.

At the \textbf{else if} statement we know that  $P[1] > Sh[1]$.
Hence, if there is only one element remaining in $Sh$ then,
by $A1$, $A2$ and $A5$, that element should be inserted between the end
of $S$ and the beginning of $P$, at which point $S$ is ordered.
That is accomplished by the right rotation by one.
The length of $P$ is unchanged but shifted one to the right, which preserves $A0$.
Setting $i = i+1$ and $j=j+1$ shifts $P$ empties $Sh$ which terminates the loop.

At entry to the last \textbf{else} section we have that $|Sh| \geq 2$ so that the scan procedure can
function and we have the general situation illustrated in Figure \ref{sublist1}.
The Unshuffle, which is an Un-in-shuffle, creates $E$, all the even indexed elements of $D$, 
and $O$, the odd indexed elements.
By $A5$, all the elements of $E$ are of the same type as $P$ and, by $A4$, greater than any element of $P$.
All the elements in $O$ are, by Scan, less than $P[1]$ and, by $A1$, all greater than any element of $S$.
Hence the rotation and redefinition of the lists preserves $A1$ - $A5$.
$P$ loses only one element, $P[1]$, and gains at least one from $E$, which preserves $A0$.

At each traversal of the loop $i$ is incremented by at least one, increasing the length of $S$.
But $P$ is never empty.
Hence at some point $Sh$ must become empty and the process terminates.

\subsection{The General Case}

Algorithm 1 just described, as in \cite{Dalkilic:2013}, will only handle lists of equal length.
As first described in \cite{EllMar00},
to handle the general case of unequal lengths we use a mirror image of the \emph{right-going-merge}
called the \emph{left-going-merge}.
The left-going-merge scans from right to left, with appropriately modified element comparisons, and
the rotate shifts to the left by $r$.

\begin{figure}
\begin{center}
\epsfig{file=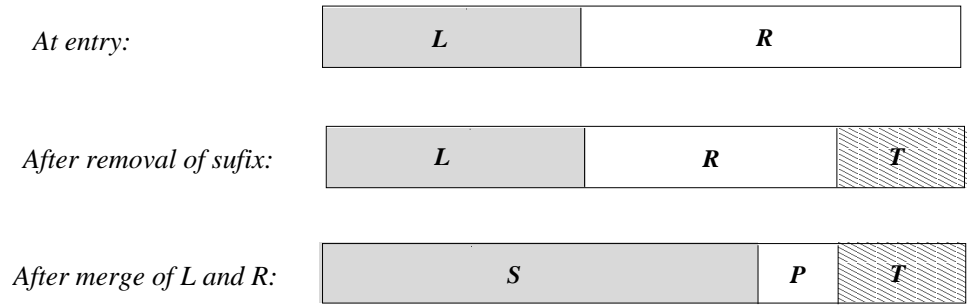, width=5in}
\end{center}
\caption{The case of unequal length lists}
\label{sublist3}
\end{figure}

If the input lists are of unequal lengths, we remove a prefix or a suffix, $T$, from the longer list
to make the lengths equal.
Then we invoke either the left or right going merges depending on whether $T$ was a prefix or a suffix.
At end of the merge, $P$ may not be empty, in which case it may need to be merged into $T$.
But that is only necessary if $P$ and $T$ are of opposite types.
The process is illustrated in Figure \ref{sublist3} and described in Algorithm 2.

The variable \emph{type} has just two possible values, \emph{left} and \emph{right}.
The value of \emph{type} indicates the type of the elements in $P$, i.e., whether they were
originally from the left or right list.
The value of \emph{type} is initialised by the outer layer of the algorithm. See Algorithm 2.
It can change during the merge procedures but the only place where it changes is where
$|P|$ is decreased to zero. 
At that point, see Algorithm 1, $type$ is complemented.

\subsection{Correctness of the general case}
The correctness of the general case follows from the correctness of the case where the lists are of equal
length and the following observations.
The loop in Algorithm 2 sets the value of \emph{type} correctly, depending on which merge is going to be invoked.
Because the loop condition ensures that $P$ is not empty at the termination of either merge procedure,
$P$ is not empty at the return from those procedures in the loop.
Hence the value of \emph{type} indicates whether the elements in $P$ at this point were originally from the left or 
right input lists.  
It follows that, if $P$ and $T$ are of the same type then by $A0$, $A2$ and the fact that, by construction,
all elements of $T$ are greater than any element in $P$, then the entire array is ordered.
Otherwise nothing is known about the relative values of the elements in $P$ and $T$ and so the process is repeated with
$P$ and $T$ as the input lists.

\begin{algorithm}
\begin{tabbing}
1234\=1234\=1234\=1234\=12345\=12345\=12345\=12345         \kill
\>\{Let $L$ and $R$ denote the input left and right lists, respectively\}\\
\>complete := false;\\

\>\textbf{while} \textbf{not} complete \textbf{do}\\

\>\>\textbf{if} $|R| > |L|$\\
\>\>\textbf{then} \{right list is longer than left list\}\\
\>\>\>remove a suffix, say $T$, from $R$ so that $|L| = |R|$;\\
\>\>\>type := left; apply the right-going-merge to $L$ and $R$;\\
\>\>\>\textbf{if} type = left \textbf{then} redefine $L$ as $P$ and $R$ as $T$ \textbf{else} complete := true \textbf{fi}\\
  \\
\>\>\textbf{else if} \{left list is longer than right list\}\\
\>\>\textbf{then} remove a prefix, say $T$, from $L$ so that $|L| = |R|$;\\
\>\>\>type := right; apply the left-going-merge to $L$ and $R$;\\
\>\>\>\textbf{if} type = right \textbf{then} redefine $L$ as $T$ and $R$ as $P$ \textbf{else} complete := true \textbf{fi}\\
\\
\>\>\textbf{else} \{the lists are of equal length\}\\
\>\>\>type := left; apply the right-going-merge to $L$ and $R$;  complete := true \textbf{fi}\\
\>\>\textbf{fi}\\
\>\textbf{endwhile}\\
\end{tabbing}
\caption{The outer layer of the algorithm}
\end{algorithm}

\newpage
\section{Time Complexity}
First consider the procedures left-going-merge and right-going-merge.

\subsection{A worst case}

There is an $\Omega(n^2)$ worst case.
Suppose that, during the first pass through the inner loop, scan finds $D$ to be of length $N/2$.
Then the unshuffle leaves $|P|$ to be of length $N/4$.
Then suppose that $N/4$ of the odd indexed elements
in $Sh$ of the opposite type to $P$ need to be interspersed evenly among the elements of $P$.
For example, suppose $P$ is a sequence of even integers and the odd elements in $Sh$ are the
missing odd integers.
This requires rotating $P$, of average length $N/8$, $N/4$ times, i.e., $\Omega(n^2)$ time.

\subsection{The average case}

The standard definition of average or expected time on inputs of a given size is the total of
all execution times over all instances of that size divided by the number of instances of that size,
i.e., we consider all instances to be equally likely.
We show that, despite the existence of a worst case, the average case time complexity is linear.
In Lemmas 3.3, 3.4 and 3.5, each of the procedures, Scan, Unshuffle and Rotate, is shown to take 
constant time on the average.
This is because, although the execution time is proportional to the length of the input, the 
probability that the input is of a certain length decreases exponentially with that length.

During the Scan procedure  the first element in $P$, $P[1]$, is compared to odd
elements in $Sh$ of opposite type until an element greater than that $P[1]$ is found, or until $Sh$ is exhausted.
We denote the prefix of $Sh$ thus discovered by $D$ and its length by $2r$, since its length
is even by definition.

\begin{lemma}
\label{Prscan}
If $r>0$ then $Pr(|D| = 2r) \leq 1/2^r$. 
\end{lemma}
\noindent
\textbf{Proof}
\ \ \\
Let the number of elements in $P$, which are all from one list, say $L$, plus the number of $L$
elements in $Sh$, be $n$ and the number of $R$ elements in $Sh$ be $m$.
The number of possible merged arrangements of these $n+m$ elements is $(n+m)!/(n! m!)$.

Suppose Scan defines $D$ such that $|D| = 2r > 0$.
We note that $m\geq r$ and $n \geq m$.
After the Unshuffle, Rotate and redefinition of the list, the number of $L$ elements in $P$ and 
$Sh$ is $n-1$ and the number of $R$ elements in $Sh$ is $m-r$.
See Figure \ref{sublist1}.
The number of arrangements consistent with this fact is $(n+m-r-1)! / ((n-1)!(m-r)!)$.
Hence the probability that  $|D| = 2r$ is given by:

\begin{eqnarray}
Pr(|D| = 2r) &=& \frac{(n+m-r-1)! n! m!} {(m-r)! (n-1)! (n+m)!} \nonumber \\
      &=& \frac{n.m(m-1) \cdots (m-r+1)}{(n+m)(n+m-1) \cdots (n+m-r)} \nonumber \\ 
      &=& \frac{n}{n+m-r} \times \frac{m}{n+m} \times \frac{m-1}{n+m-1} \cdots \times \frac{m-r+1}{n+m-r+1} \nonumber \\
      & \leq & 1/2^r \nonumber
\end{eqnarray}

because $m \geq r$ implies $n/(n+m-r) \leq 1$ and, for all $0 \leq k $, $(m-k)/(n+m-k) \leq 1/2$.
\enproof

Let $P_b$ and $S_b$ be the $P$ and $S$ lists, respectively, at the bottom of the \textbf{while} loop,
after all rearrangements and adjustments to the lists.
See Figure \ref{sublist1}.

\begin{lemma}
\label{PrlenP}
$Pr(|P_b| = p) \leq  1/2^{p-1}$.
\end{lemma}
\noindent
\textbf{Proof}
\ \ \\
Suppose there are $n$ elements of type \emph{left} and $m$ of type \emph{right} distributed across $S_b$ 
and $P_b$.
Then $m$ is within one of $n$ because the numbers of each type remaining in $Sh$ are within one
of each other and the input lists were of equal length.
The number of possible merged arrangements of these $m+n$ elements is $(m+n)!/(m!n!)$.

Without loss of generality, suppose the elements in $P_b$ are of type \emph{left}.
Then the number of arrangements consistent with the existence of $|P_b| = p$ elements all greater than any
element in $S_b$ is the number of ways that $S_b$ can result from the merge of $n$ with $m-p$
elements, i.e.,  $(m+n-p)!/((m-p)! n!)$.
Hence the probability that $|P_b| = p$ is given by:
\begin{eqnarray}
Pr(|P_b| = p) &=& \frac{(m+n-p)!} {(m-p)! n!} \times \frac{m! n!}{(m+n)!} \nonumber \\
      &=& \frac{ m(m-1)(m-2)\cdots (m-p+1)} {(m+n)(m+n-1)(m+n-2) \cdots (m+n-p+1)} \nonumber \\
      &=& \frac{m}{(m+n)} \times \frac{m-1}{m+n-1} \times \frac{m-2}{m+n-2} \cdots \times \frac{m-p+1}{m+n-p+1} 
           \nonumber \\
      & \leq & 1/2^{p-1} \nonumber
\end{eqnarray}
because $m/(m+n) < 1$ and, for all $0 \leq k < n$, $(m-k)/(m+n-k) \leq 1/2$. 
\enproof

Lemmas \ref{Prscan} and \ref{PrlenP} allow us to show that the expected time complexity of all the 
loop procedures is a constant.

\begin{lemma}
\label{tscan}
The expected time used by the Scan procedure is constant.
\end{lemma}
\noindent
\textbf{Proof}
\ \ \\
The time taken to scan $2r$ elements is $k_1r$, for some constant $k_1$.
Hence the expected time, \emph{expt-scan}, is given by:
\begin{eqnarray}
\emph{expt-scan} &=& \sum_r k_1rPr(|D|=2r) \leq \sum_{r=1}^\infty k_1r/2^r = 2k_1, \nonumber
\end{eqnarray}
by Lemma \ref{Prscan}.
\enproof

\begin{lemma}
\label{tunshuff}
The expected time used by the unShuffle procedure is constant.
\end{lemma}
\noindent
\textbf{Proof}
\ \ \\
The time required to Unshuffle a list $D$, where $|D| = 2r$, is $k_2r$, for some constant $k_2$.
See Section 4 below.
Hence the expected time, \emph{expt-shuff}, is given by:
\begin{eqnarray}
\emph{expt-shuff} &=& \sum_r k_2rPr(|D|=2r) \leq \sum_{r=1}^\infty k_2r/2^r = 2k_2, \nonumber
\end{eqnarray}
by Lemma \ref{Prscan} and the fact that the unshuffle works on the result of the scan.
\enproof

\begin{lemma}
\label{trotate}
The expected time used by the Rotate procedure is constant.
\end{lemma}
\noindent
\textbf{Proof}
\ \ \\
$P_b+1$ elements are rotated, where $P_b$ is list $P$ at the bottom of
the loop, i.e., after rotation.
The time taken to rotate $|P_b|+1$ elements is $\leq (|P_b|+1) k_3$, for some constant $k_3$.
Hence the expected time, \emph{expt-rot}, is given by:
\begin{eqnarray}
\emph{expt-rot} &=& \sum_p (p+1)k_3Pr(|P_b| = p)\leq \sum_{p=1}^\infty (p+1)k_3/2^p = 
    \sum_{p=1}^\infty pk_3/2^p +  \sum_{p=1}^\infty k_3/2^p = 3k_3, \nonumber
\end{eqnarray}
by Lemma \ref{PrlenP}.
\enproof

\begin{theorem}
\label{avgtime}
The average time complexity of the algorithm is $O(n)$, where $n$ is the combined length of the
lists.
\end{theorem}
\noindent
\textbf{Proof}
\ \ \\
Consider the merge procedures.
The actions inside the loop are either constant time operations or scans, rotations or unshuffles
which, by Lemmas \ref{tscan}, \ref{tunshuff}, \ref{trotate} and the ``linearity of expectations''
\cite[Appendix C]{cormen:alg:2001}, are expected constant time operations.
Hence the expected time to traverse the \textbf{while} loop is a constant.

Now consider the general case where the problem is broken down to the merge of a sequence of equal length
lists, say $n_1$, $n_2 \ldots  n_k$.
We observe that $n_1 \geq n_2 \ldots \geq n_k$ and that \(\sum_{i=1}^k n_i = n\).
Since each merge takes time $O(n_i)$, the total time is $O(n)$.
\enproof

\section{Shuffling}
An algorithm that performs the perfect shuffle, in-place, is not obvious.
In the original paper \cite{EllMar00}, amongst other suggestions, it was pointed out that a simple cycle leader
algorithm that used one bit of extra memory per element would provide a simple solution.
The extra bits are used to mark elements already moved and permit finding an unused element,
i.e., a new leader, from which to start a new cycle when the current cycle ends.
Although one bit per element violates the $O(\log n^2)$ space restriction, it might be acceptable in practice.

We cite two papers that suggest practical solutions that function within the desirable constraints,
i.e., $O(\log^2 n)$ space and linear time.
In \cite{Jain2008} a cycle leader algorithm is given in which some simple number theory is used to
generate the ``leaders''.
It uses linear time and absolutely minimum space i.e., $O(\log n)$.

In \cite{YangEMR13} it is shown that the shuffle can be attained by a sequence of 
element exchanges where no element participates in more than two exchanges.
A couple of ways of computing the identity of the elements to be exchanged are provided.
One of these methods uses linear time and $O(\log^2 n)$ space.

\section{Stability}
A useful feature of using the perfect shuffle is that stability falls out without much extra effort.
We note that neither the shuffling, unshuffling nor the rotations, which are the only processes that move
elements, can change the order of any pair of elements that are originally from the same list.
Comparisons between elements from different lists are done in the scan procedure and at the first \textbf{if}
in the loop.
So it only remains to be sure that those comparisons give the correct priority when comparing equal elements.
We note that comparisons are always between the first element in $P$ and an odd indexed element in $Sh$
which are, by $A3$ and $A5$, of opposite types.
So assigning correct priority is easily done by using the current value of \emph{type}
which tells us the current type of the elements in $P$.

\section{Conclusions}
We have given an analysis of the time complexity of the improved merge via shuffling algorithm presented originally 
in \cite{Dalkilic:2013} where only experimental timing results on special cases were presented.
We have shown that, although there is an $\Theta(n^2)$ worst case, the average case is linear time.
We provide a more concise definition of the algorithm which allows a proof of correctness.
We have expanded the description of the algorithm to include the general case, where it was originally restricted to
the case of equal length lists and we have included a mechanism that ensures stability, not given in the original.
We have cited one more recent shuffling method, not mentioned in the original paper.

We suggest that, by using the recently described Shuffling techniques cited in Section 5, 
this algorithm may provide a practical alternative to data processes willing to trade a slower execution time 
for a halving of internal memory usage.
\bibliography{sort-merge}
\bibliographystyle{alpha}
\end{document}